\documentclass[10pt,twocolumn,letterpaper]{article}

\usepackage[pagenumbers]{cvpr}

\usepackage{graphicx}
\usepackage{amsmath}
\usepackage{amssymb}
\usepackage{booktabs}
\usepackage{multirow}
\usepackage{enumitem}
\usepackage[table]{xcolor}

\usepackage[pagebackref,breaklinks,colorlinks]{hyperref}
\usepackage{mathrsfs}

\usepackage[capitalize]{cleveref}
\crefname{section}{Sec.}{Secs.}
\Crefname{section}{Section}{Sections}
\Crefname{table}{Table}{Tables}
\crefname{table}{Tab.}{Tabs.}

\begin{document}

\title{ColDE: A Depth Estimation Framework for Colonoscopy Reconstruction}

\author{
Yubo Zhang\textsuperscript{1} \quad
Jan-Michael Frahm\textsuperscript{1} \quad
Samuel Ehrenstein\textsuperscript{1} \quad
Sarah K. McGill\textsuperscript{2} \quad\\
Julian G. Rosenman\textsuperscript{3} \quad
Shuxian Wang\textsuperscript{1} \quad 
Stephen M. Pizer\textsuperscript{1} \vspace{3pt}\\  
\textsuperscript{1}Department of Computer Science, University of North Carolina at Chapel Hill \\
\textsuperscript{2}Department of Medicine, University of North Carolina at Chapel Hill\\ 
\textsuperscript{3}Department of Radiation Oncology, University of North Carolina at Chapel Hill \\ 
{\tt\small \{zhangyb,jmf,ehrensam,shuxian,pizer\}@cs.unc.edu} \\
{\tt\small mcgills@email.unc.edu} \quad
{\tt\small rosenmju@med.unc.edu}
}
\maketitle

%%%%%%%%% ABSTRACT
\begin{abstract}
One of the key elements of reconstructing a 3D mesh from a monocular video is generating every frame's depth map.
However, in the application of colonoscopy video reconstruction, producing good-quality depth estimation is challenging.
Neural networks can be easily fooled by photometric distractions or fail to capture the complex shape of the colon surface, predicting defective shapes that result in broken meshes.
Aiming to fundamentally improve the depth estimation quality for colonoscopy 3D reconstruction,
in this work we have designed a set of training losses to deal with the special challenges of colonoscopy data.
For better training, a set of geometric consistency objectives was developed, using both depth and surface normal information.
Also, the classic photometric loss was extended with feature matching to compensate for illumination noise.
With the training losses powerful enough, our self-supervised framework named \textbf{ColDE} is able to produce better depth maps of colonoscopy data as compared to the previous work utilizing prior depth knowledge.
Used in reconstruction, our network is able to reconstruct good-quality colon meshes in real-time without any post-processing, making it the first to be clinically applicable.
\end{abstract}

%%%%%%%%% BODY TEXT
\section{Introduction}
\begin{figure}[t]
    \centering
    \includegraphics[width=0.47\textwidth]{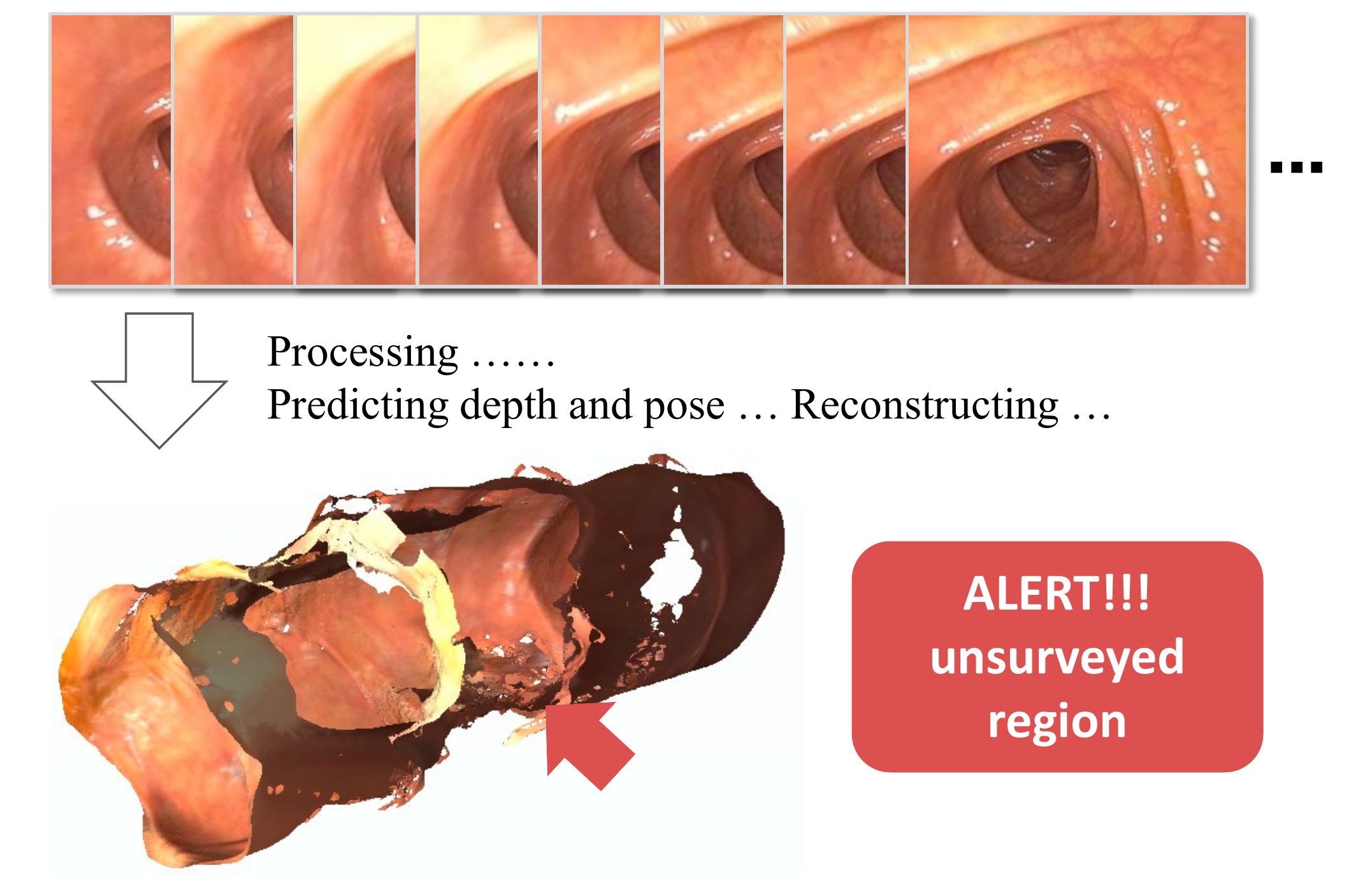}
    \caption{Reconstructing the 3D mesh from a colonoscopy video in real-time according to the predicted depth and camera pose, allowing holes in the mesh to alert the physician to unsurveyed regions on the colon surface.}
    \label{fig:intro}
    \vspace{-5pt}
\end{figure}

Colon cancer being one of the deadliest cancers worldwide, optical colonoscopy has been the most efficient for pre-cancerous screening.
The procedure is operated by a physician, who controls a colonoscope inserted inside the patient's large intestine (colon) and detects and removes lesions (polyps) that are visible on the live RGB video sent back from the scope.
To increase the polyp detection rate, it is crucial to lowering the percentage of the colon surface that is missed from examination~\cite{hong2007colonoscopy}.
As the unsurveyed regions are usually caused by the lack of camera orientations or occlusion by the colon structure itself,
our approach is to reconstruct the already surveyed region from the colonoscopy video concurrently during the procedure so that the unsurveyed part can be reported to the physician as holes in the 3D surface (as in Fig.~\ref{fig:intro}).

To reconstruct the 3D model from the monocular video, dense depth and camera pose transformation need to be predicted for each frame.
Previous deep learning methods have been developed for that task in daily applications such as outdoor driving~\cite{Geiger2012CVPR,cordts2016cityscapes} and scene reconstruction~\cite{saxena2008make3d,li2018megadepth}.
In most scenarios where ground-truth information is not available, photometric objectives have been proposed to train the network in a self-supervised fashion~\cite{zhou2017unsupervised,zhan2018unsupervised}.
Later, additional geometric constraints have been explored~\cite{bian2019unsupervised,yang2018lego,mahjourian2018unsupervised}, and other visual clues such as optical flow~\cite{yin2018geonet,ranjan2019competitive} and segmentation~\cite{casser2019depth,guizilini2019semantically} have also been utilized to improve the performance.
However, these approaches were primarily developed to estimate in the human-made world.
Their validity to cope with colonoscopy data has not been fully studied, where the sparse texture and the poor illumination bring a great challenge to generate estimation, especially depth maps.

Due to the difficulty in estimation of the depth map directly from optical colonoscopy, previous work that dealt with the problem often used the help from prior knowledge available in virtual colonoscopy~\cite{mathew2020augmenting} or colon simulators~\cite{zhang2020template,rau2019implicit}.
The first framework~\cite{ma2019real,ma2021rnnslam} that successfully reconstructed 3D models from real colonoscopy videos generated sparse pseudo ground-truth using structure-from-motion (SfM)~\cite{schonberger2016structure} so that the depth estimation network could be trained with semi-supervision.
But even with the help of prior knowledge, the depth prediction was still vulnerable to environmental noise and failed to capture the surface information from time to time.
When the depth predictions of frames were placed together to generate a 3D reconstruction mesh, their shapes were not aligned, causing a sparse and broken surface.
To compensate, the authors introduced an additional averaging step and adjustment of the depth using SfM~\cite{bae2020deep,liu2020reconstructing}.
However, these post-processing steps prevented the programs' real-time execution.

% ColDE
In this work we introduce \textbf{ColDE}, a novel framework for \textbf{Col}onoscopy \textbf{D}epth \textbf{E}stimation, where we fundamentally improve the depth prediction quality for colonoscopy 3D reconstruction. Our contributions are the following:
\begin{itemize}[leftmargin=*]
\item We have designed our training losses to deal with the special challenges of colonoscopy data.
A set of geometric consistency losses, constraining the surface's 3D positions and normals, has been developed to enforce the shape consistency between frames.
Also, the photometric loss has been extended with image feature matching, improving the network's robustness to illumination change.
\item With these powerful training objectives, our framework is the first to be able to train the network in a fully self-supervised fashion from real colonoscopy videos without needing any prior knowledge.
\item When tested on the colon simulator data~\cite{zhang2020template}, our framework outperforms previous state-of-the-art monocular depth estimation methods on depth errors and accuracy.
\item 
Applied to optical colonoscopy data, our 3D reconstruction pipeline is the first to be clinically applicable,
because it is able to generate depth maps in real-time that lead to meshes with desirable quality.
\end{itemize}

\section{Related Work}
Here, we review the previous methods of estimating a depth map from a single RGB image
and their use in reconstructing endoscopy videos.

\subsection{Monocular Depth Estimation}
Estimating the depth map of a single RGB image is an ill-posed problem.
As deep learning has progressed, various networks and training methods have been developed to address the problem.
When the ground-truth depth information is fully available, supervised training methods can lead to decent depth prediction~\cite{eigen2015predicting,vijayanarasimhan2017sfm,fu2018deep}.
Others have explored the possibility to utilize sparse depth values obtained from sensors or other odometry approaches~\cite{kuznietsov2017semi,yang2018deep}.

However, acquiring the ground-truth depth information is challenging, so applying unsupervised training strategies is more applicable in some real-world scenarios.
Godard et al. \cite{godard2017unsupervised} proposed one of the earliest self-supervised approaches to generate depth maps from stereo images;
then Zhou et al. \cite{zhou2017unsupervised} extended the method and developed the first self-supervised monocular depth estimation framework.
Later work explored various directions to improve the performance on outdoor~\cite{Geiger2012CVPR,cordts2016cityscapes} and indoor~\cite{silberman2012indoor} depth estimation tasks.
Some focused on solving specific issues that often occur in training and at test time, e.g., the scale inconsistency issue~\cite{wang2018learning,xiong2020self,ji2021monoindoor} and the mixed information brought in by independent motion or static frames~\cite{monodepth2,bian2019unsupervised,li2020unsupervised}.
Better network architectures have been proposed \cite{packnet,zou2020learning,lyu2020hr,you2021towards}.
Additional training objectives that extend the photometric constraint~\cite{zhan2018unsupervised,yang2018unsupervised} or introduce geometric constraints~\cite{yang2018unsupervised,yang2018lego,mahjourian2018unsupervised,bian2019unsupervised,li2021structdepth} have been found useful.
Other visual information, such as optical flow~\cite{yin2018geonet,zou2018df,ranjan2019competitive} and semantic segmentation~\cite{casser2019depth,meng2019signet,guizilini2019semantically}, were also utilized to help the depth estimation.

\subsection{3D Reconstruction in Endoscopy}
Applying monocular depth estimation to endoscopy data, such as optical colonoscopy, is challenging due to the low-texture surface and the complex lighting.
As pure self-supervised methods may struggle to complete the task, previous work tried to augment the training data with ground-truth depth information, which is only fully available in virtual colonoscopy.
In order to transfer the knowledge learned from synthetic data to real data,
several methods~\cite{mahmood2018unsupervised,mathew2020augmenting,cheng2021depth} made use of GAN frameworks and Itoh et al. \cite{itoh2021unsupervised} designed a reflection model.

In the work by Ma et al. \cite{ma2019real,ma2021rnnslam}, with the help of structure-from-motion (SfM)~\cite{schonberger2016structure} to generate sparse depth information, the depth estimation network~\cite{wang2019recurrent} was trained in a semi-supervised fashion with only real data.
Then accompanied by a SLAM system~\cite{engel2017direct} to predict camera poses, 
they developed the first framework that was able to reconstruct colon meshes from colonoscopy videos in real-time.
However, their depth estimation network can only handle simple cases and is vulnerable to environmental noise, and the predicted shapes 
often fail to produce good quality meshes.
Later work~\cite{bae2020deep,liu2020reconstructing} exploited the possibility to integrate SfM with the learning-based depth estimation to calibrate depth predictions, 
but the time expense brought in by SfM restricts the methods from large-scale reconstruction applications.

To improve these endoscopy reconstruction prototypes, in this work we target their weak link, which is the neural depth estimation component.
Using only the optical (RGB) colonoscopy data available, we aim to substantially improve the effectiveness and efficiency of the reconstruction by a better training strategy.
\begin{figure*}[!ht]
    \centering
    \includegraphics[width=0.98\textwidth]{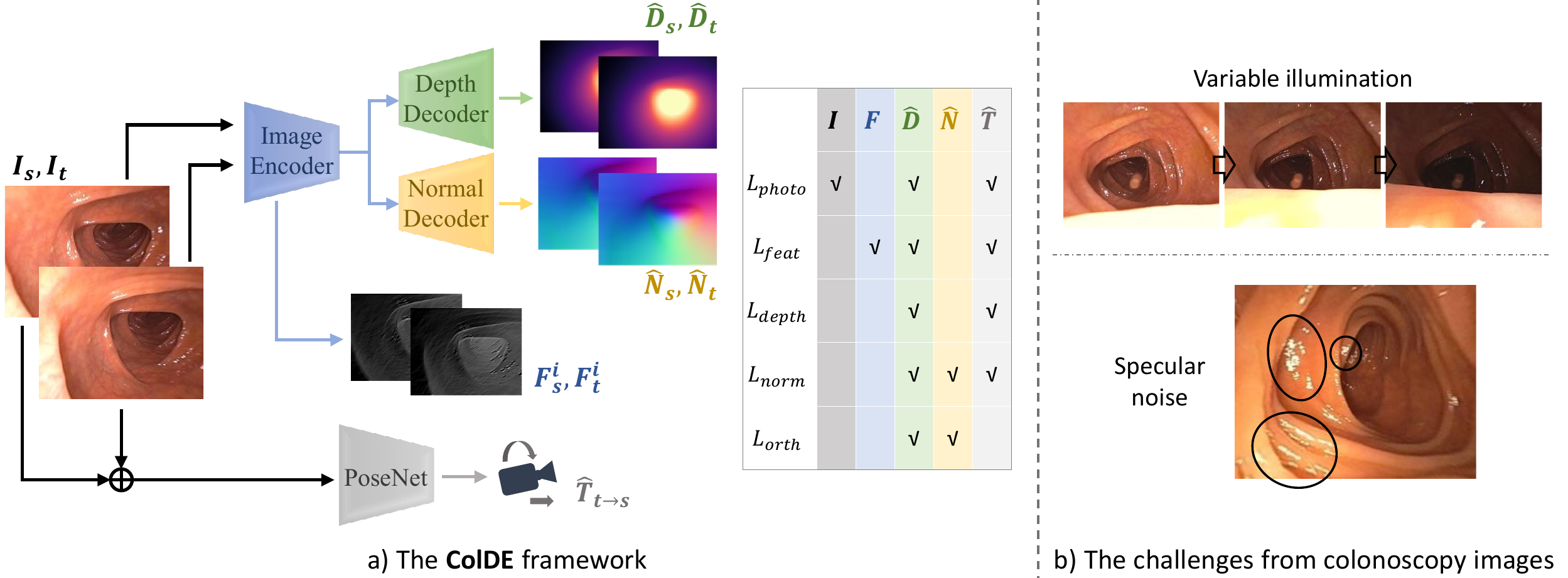}
    \caption{a) Overview of our \textbf{ColDE} framework. Images are encoded by an image encoder, then the depth and normal map are generated by their individual decoders. A separate PoseNet takes a pair of images and predicts their relative camera transformation. The relations of loss functions and their conditional variables are listed in the table on the right.
    b) The photometric noise in colonoscopy images, such as changing contrast or specular noise, can be challenging for the network to deal with.}
    \label{fig:network}
    \vspace{-5pt}
\end{figure*}

\section{Methods}
In most of the previous monocular depth estimation frameworks~\cite{zhou2017unsupervised,monodepth2,bian2019unsupervised}, the self-supervision signals that train the network come from the projection relation between a source view $s$ and a target view $t$.
Given the camera intrinsic $K$, a pixel $p_t$ in a target view can be projected into the source view according to the predicted depth map $\hat D_t$ and the relative camera transformation $\hat T_{t \to s}$.
This process yields the pixel's homogeneous coordinates $\hat p_s$ and its projected depth $\hat d_s^t$ in the source view, as in Eq.~\ref{Eq.proj}:
\begin{gather}
    \label{Eq.proj}
    \hat p_s, \hat d^t_s \sim K \hat T_{t \to s} \hat D_t(p_t) K^{-1} p_t
\end{gather}

Here, we describe our colonoscopy depth estimation framework shown in Fig.~\ref{fig:network} a.
Building upon the projection relation in Eq.~\ref{Eq.proj}, we construct a set of self-supervised photometric and geometric objectives, which target the colonoscopy video reconstruction task that is challenging due to lighting complexity and complicated topology.
These objectives are used to train the depth and pose estimation networks; details will be discussed later in this section.

\subsection{Photometric Objectives}
The first set of training objectives is based on the photometric information.
We will first discuss using the RGB image appearance as the objective and then using feature maps.

\subsubsection{Appearance Similarity Objective}
Based on the projection in Eq.~\ref{Eq.proj}, the source image $I_s$ can be warped into the target view $I_t$ as in Eq.~\ref{Eq.warp}, where $\left \langle * \right \rangle$ denotes bilinear interpolation.
Assuming the scene is rigid and the light source does not change much, which is the case for most of a colonoscopy video, the warped source image $\hat I_s$ should agree with the target image,
so with the appearance difference we formulate the photometric similarity objective $L_{photo}$ to supervise the depth and camera pose training, as in Eq.~\ref{Eq.photo}. 
$\rm SSIM(*,*)$ denotes the structural similarity measurement~\cite{wang2004image}.
Following \cite{godard2017unsupervised,monodepth2} we set $\alpha = 0.85$.
\begin{gather}
    \label{Eq.warp}
    \hat I_s = I_s \left \langle \hat p_s \right \rangle \\
    \label{Eq.photo}
    L_{photo} = (1 - \alpha) || \hat I_s - I_t ||_1 + \frac{\alpha}{2} (1 - {\rm SSIM}(\hat I_s, I_t))
\end{gather}

\subsubsection{Feature Similarity Objective}
Trained with only the appearance clue (image intensity), the neural network can be vulnerable to photometric variation in the environment.
In colonoscopy videos the light source is moving with the camera and obstacles often occur to block the lumination, 
leading to changing contrast and brightness~\cite{zhang2021lighting}.
Moreover, the watery surface can reflect the light and cause notable specular noise,
as shown in Fig.~\ref{fig:network} b.

To deal with this issue, in addition to the appearance-based photometric loss, we have designed a feature-based similarity objective~\cite{zhan2018unsupervised}.
We utilize the early features from the depth prediction network (conv$1$ output of ResNet encoder~\cite{he2016deep}), which represent local semantic information.
In each training iteration, we randomly sample one feature map out of $64$ to be used as the source and target ``images'', $F^i_s$ and $F^i_t$,
and the feature similarity objective $L_{feat}$ is constructed with the target and warped source feature maps as
\begin{gather}
    L_{feat} = \frac{\alpha}{2} (1 - {\rm SSIM}(F^i_s \left \langle \hat p_s \right \rangle, F^i_t)), i \ {\rm in} \left\{ 1, ..., 64 \right\}
\end{gather}
Here we only include the semantics-communicating structural similarity term in feature matching.
Additionally, during training $L_{feat}$ directly supervises the first step of feature extraction, further making the process robust to variable lighting.

\subsection{Geometric Objectives}
The network trained with photometric objectives may seem to produce decent depth maps for individual images.
But when they are stitched together to produce the overall 3D model of the scene, the projected shapes of adjacent frames may not align with each other, causing defective reconstruction~\cite{ma2021rnnslam}.
To cope with this issue, we design another set of training objectives to enforce geometric consistency,
specifically, on 3D positions and surface normals.

\subsubsection{Depth Consistency Objective}
Directly measuring the geometric consistency between two shapes is not easy~\cite{mahjourian2018unsupervised}, so instead we measure the difference between the predicted depths of the same scene in different frames.
According to Eq.~\ref{Eq.proj}, each vertex in the predicted 3D shape of the target frame (derived from $\hat D_t$) has the depth of $\hat d^t_s$ relative to the source view's camera.
Meanwhile, the network can also produce a depth map $\hat D_s$ for the source image.
As the predicted depth of the source image should agree with the projected depth from the target image, we can construct a depth consistency objective as in Eq.~\ref{Eq.depth}, in which we adopt the means in~\cite{bian2019unsupervised} to normalize the depth:
\begin{gather}
\label{Eq.depth}
    L_{depth} = \frac{\left| \hat D_s \left \langle \hat p_s \right \rangle - \hat D^t_s \right|}{\hat D_s \left \langle \hat p_s \right \rangle + \hat D^t_s}
\end{gather}
With the depth consistency enforced, when projecting images into the 3D space, the 3D models from the target and source views will be better aligned and can be stitched together closely.

\subsubsection{Normal Consistency Objective}
Surface normals provide useful information about objects' shape~\cite{yang2018unsupervised}.
It describes the orientation of the 3D surface and reflects local shape knowledge.
As the derivative of vertices' 3D positions, they can be sensitive to the error and noise on the predicted surface.
Therefore, utilizing surface normal correspondence during training can further correct the network prediction and improve the shape consistency.

Let $\hat N_t$ be the object's surface normals in the target coordinate system.
Then in the source view's coordinate system those vectors will be in different directions depending on the relative camera rotation $\hat R_{t \to s}$ (the rotation component of $\hat T_{t \to s}$).
These directions from the target view should agree with the source view's own normal prediction $\hat N_s$; using this correspondence we form the normal consistency objective as
\begin{gather}
\label{Eq.normal}
    L_{norm} = || \hat N_s \left \langle \hat p_s \right \rangle - \hat R_{t \to s} \hat N_t ||_1
\end{gather}
Here, we simply use the numerical difference between the two vectors to compute the normal error.
In practice, we find that using angular difference has similar performance.

\paragraph{Surface Normal Prediction}
Previous practice often obtained normal vectors by computing the mean cross product of surface vector prediction~\cite{yang2018unsupervised,yang2018lego}.
However, we found that when training with colonoscopy data, computing normals directly from depths is less stable and tends to result in unrealistic shapes, since the colon surface is less flat compared to the surfaces in human-made worlds.
Instead, we apply a separate decoder to predict surface normal information.
The normal decoder has almost the same architecture as the depth decoder, only with a different number of output channels, and two decoders share the same image encoder, as shown in Fig~\ref{fig:network} a.
To bridge the depth and normal predictions, we further apply an orthogonality constraint:
\begin{gather}
\label{Eq.orth}
    \hat V(p) = \hat D(p_a) K^{-1} p_a - \hat D(p_b) K^{-1} p_b \\
    L_{orth} = \sum_p \hat N(p) \cdot \hat V(p)
\end{gather}
where $\hat V(p)$ is the approximate surface vector around $p$, which is computed from the depths of $p_a$ and $p_b$, $p$'s nearby pixels.
In practice, we apply two pairs of $p_{a/b}$ position combinations, i.e., $p$'s top-left/bottom-right pixels and top-right/bottom-left pixels.

\subsection{Training Overview}
Apart from the photometric and geometric objectives discussed above, we follow the previous work~\cite{godard2017unsupervised} and include an edge-aware smoothness prior $L_{sm}$, which is applied to both source and target views' depth predictions.
But in contrast to previous work~\cite{zhan2019self}, we have not found it useful to include a normal smoothness prior, since normal orientations of the colon surface do not necessarily follow the intensity clue in colonoscopy images.

To mask out the pixels of stationary frames in colonoscopy training sequences, we follow the auto-masking scheme in \cite{monodepth2} applying a mask $M_{auto}$.
We also follow \cite{ma2021rnnslam} to add a specular mask $M_{spec}$ to mask out specular regions.
Additionally, a valid mask $M_{valid}$ is applied to exclude the pixels that would be warped outside of the source view.
So the overall mask to select pixels in consistency objectives is the combination of three masks mentioned above, where $\odot$ denotes element-wise production:
\begin{gather}
\label{Eq.mask}
    M = M_{auto} \odot M_{spec} \odot M_{valid}
\end{gather}

Summing up all the elements, the training loss to supervise the network is written as Eq.~\ref{Eq.all-loss}, where $\lambda_1$-$\lambda_5$ are the weights of each objective.
\begin{equation}
\begin{aligned}
    L =& (L_{photo} + \lambda_1 L_{feat} + \lambda_2 L_{depth} + \lambda_3 L_{norm}) \odot M \\
    &+ \lambda_4 L_{orth} + \lambda_5 L_{sm}
\end{aligned}
\label{Eq.all-loss}
\end{equation}
\section{Experiments}
We validated our \textbf{ColDE} framework on colon simulator data and real optical colonoscopy videos.
We also tested our method on the KITTI dataset.
The results are analyzed in this section.

\subsection{Implementation Details}
We constructed our networks following \cite{monodepth2}.
Due to the time efficiency concern, our image encoder is built with ResNet18~\cite{he2016deep}, which has relatively few parameters.
As in Fig.~\ref{fig:network} a, the image encoder takes a single RGB image as the input, and the encoded image features are given to depth and normal decoders that respectively produce depth and normal maps.
The PoseNet takes a pair of images and produces their relative $6$D camera transformation.
In our experiments all input RGB images are resized to $224 \times 288$.

We used the pre-training and fine-tuning training scheme to train the network in the following experiments.
In the pre-training steps, the learning rate was set at $10^{-4}$.
First, $\lambda_{3,4}$ were set to be $0$ in Eq.~\ref{Eq.all-loss}, and we trained the image encoder, depth decoder and PoseNet with $\lambda_1=0.1$, $\lambda_2=0.1$ and $\lambda_5=0.01$ for $20$ epochs.
Then the above three modules were frozen, and we trained the normal decoder alone for $20$ epochs with $\lambda_3=0.1$ and $\lambda_4=0.5$.
Finally, all modules were fine-tuned together for $10$ epochs with learning rate $10^{-5}$.
The weights for fine-tuning were $\lambda_1=0.1$, $\lambda_2=0.1$, $\lambda_3=0.005$, $\lambda_4=0.001$ and $\lambda_5=0.01$.

\subsection{Results on Colonoscopy Simulator Dataset}
We first tested our method on the synthetic data from a colon
simulator~\cite{zhang2020template}.
In that simulator a pre-defined 3D colon model is rendered with textures similar to optical  colonoscopy videos but with less specular reflection.
As a simulated colonoscope traveling inside the colon, the RGB image and the ground-truth depth are saved for each frame.
In our experiments our data comes from $9$ simulated colonoscopy videos generated by this program, from which we randomly sampled training/validation/testing sets that contain $9.5$k/$0.5$k/$2.5$k frames.
The accuracies reported below are the results on the testing split.
 
\subsubsection{Depth Prediction Errors and Accuracy}
\begin{table*}[ht]
    \centering
    \begin{tabular}{l|cccc|ccc|c}
        \multirow{2}{*}{\bf Methods} & \multicolumn{4}{c|}{\bf Error metric $\downarrow$} & \multicolumn{3}{c|}{\bf Accuracy metric $\uparrow$} & \bf Speed $\uparrow$\\
        \cmidrule(lr){2-9}
         & Abs Rel & Sq Rel & RMSE & RMSE log & $\delta < 1.25$ & $\delta < 1.25^2$ & $\delta < 1.25^3$ & FPS\\
        \midrule
        SfMLearner~\cite{zhou2017unsupervised} & 0.276 &   0.832 & 1.592 & 0.333 & 0.660 & 0.881 & 0.947 & 26 \\
        SC-SfMLearner~\cite{bian2019unsupervised} & 0.108 & 0.132 & 0.906 & 0.171 & 0.874 & 0.957 & 0.983 & \bf 35 \\
        Monodepth2~\cite{monodepth2} & 0.097 & 0.110 & 0.811 & 0.154 & \underline{0.913} & \underline{0.971} & 0.987 & \bf 36\\
        PackNet-SfM~\cite{packnet} & 0.101 & \underline{0.106} & \underline{0.792} & \underline{0.152} & 0.901 & 0.968 & \underline{0.988} & 2 \\
        HR-Depth~\cite{lyu2020hr} & \underline{0.096} & 0.109 & 0.813 & 0.153 & \underline{0.913} & \underline{0.971} & 0.987 & 22 \\
        \midrule
        \bf ColDE & \bf 0.077 & \bf 0.079 & \bf 0.701 & \bf 0.134 & \bf 0.935 & \bf 0.975 & \bf 0.989 & \bf 36
    \end{tabular}
    \vspace{-3pt}
    \caption{Quantitative results (errors and accuracy) on colonoscopy simulator data. All results were produced from the networks trained with only self-supervised losses. The best result of each category is in bold and the second best is underlined. Our method \textbf{ColDE} outperforms previous monocular depth estimation methods on error and accuracy measures, while achieving real-time execution.}
    \label{tab:sim_sota}
\end{table*}

\begin{figure*}[!ht]
  \centering
  \resizebox{0.89\textwidth}{!}{
  \newcommand{\turnheightnew}{0.195\columnwidth}

\centering

\begin{tabular}{@{\hskip 0mm}c@{\hskip 3mm}c@{\hskip 1mm}c@{\hskip 0mm}c@{\hskip 0mm}c@{\hskip 0mm}c@{\hskip 0mm}c@{}}

 &
{\rotatebox{90}{\hspace{6mm}\scriptsize{Input}}} &
\includegraphics[height=\turnheightnew]{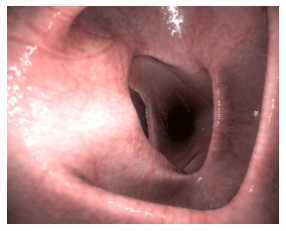} &
\includegraphics[height=\turnheightnew]{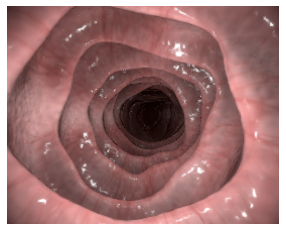} &
\includegraphics[height=\turnheightnew]{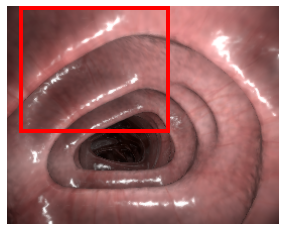} &
\includegraphics[height=\turnheightnew]{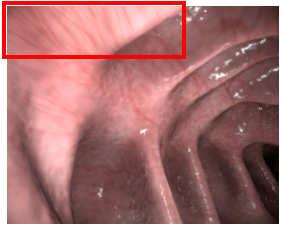} &
\includegraphics[height=\turnheightnew]{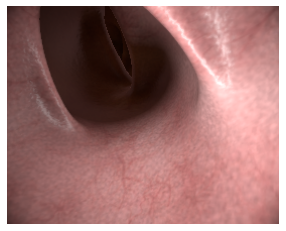}\\

 &
{\rotatebox{90}{\hspace{1.5mm}\scriptsize
{Ground-truth}}} &
\includegraphics[height=\turnheightnew]{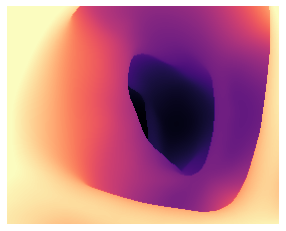} &
\includegraphics[height=\turnheightnew]{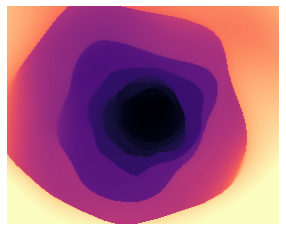} &
\includegraphics[height=\turnheightnew]{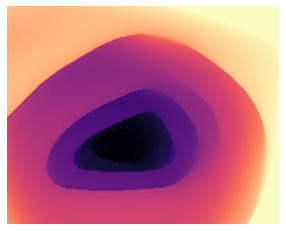} &
\includegraphics[height=\turnheightnew]{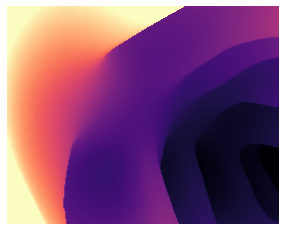} &
\includegraphics[height=\turnheightnew]{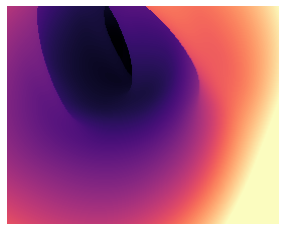}\\

\arrayrulecolor{black!30}\midrule

{\multirow{2}{*}[9.6mm]{\rotatebox{90}{\footnotesize
{Non-real-time methods}}}} &
{\rotatebox{90}{\hspace{1.5mm}\tiny
{PackNet-SfM~\cite{packnet}}}} &
\includegraphics[height=\turnheightnew]{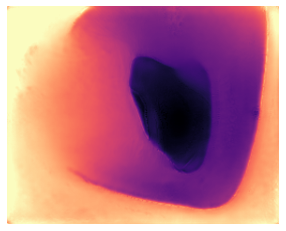} &
\includegraphics[height=\turnheightnew]{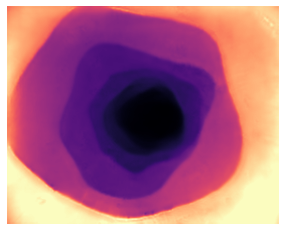} &
\includegraphics[height=\turnheightnew]{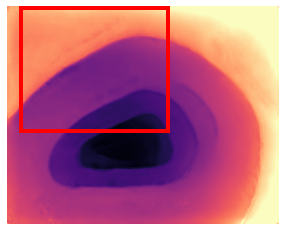} &
\includegraphics[height=\turnheightnew]{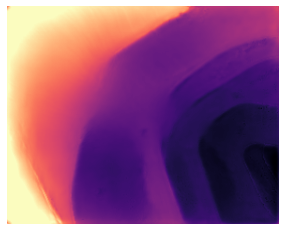} &
\includegraphics[height=\turnheightnew]{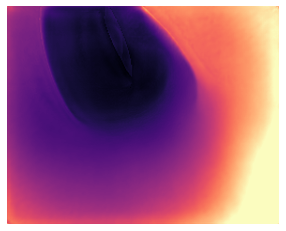}\\
 &
{\rotatebox{90}{\hspace{2.5mm}\tiny
{HR-Depth~\cite{lyu2020hr}}}} &
\includegraphics[height=\turnheightnew]{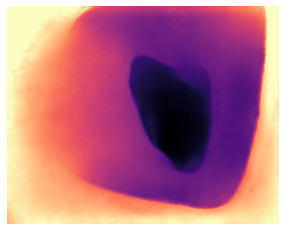} &
\includegraphics[height=\turnheightnew]{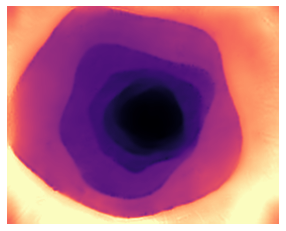} &
\includegraphics[height=\turnheightnew]{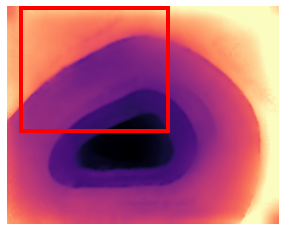} &
\includegraphics[height=\turnheightnew]{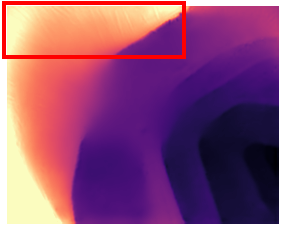} &
\includegraphics[height=\turnheightnew]{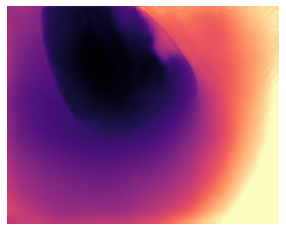}\\

\arrayrulecolor{black!30}\midrule

{\multirow{2}{*}[7mm]{\rotatebox{90}{\footnotesize
{Real-time methods}}}} &
{\rotatebox{90}{\hspace{2mm}\tiny
{Monodepth2~\cite{monodepth2}}}} &
\includegraphics[height=\turnheightnew]{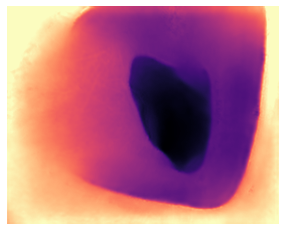} &
\includegraphics[height=\turnheightnew]{figs/sim_sota/NLY812_mono.png} &
\includegraphics[height=\turnheightnew]{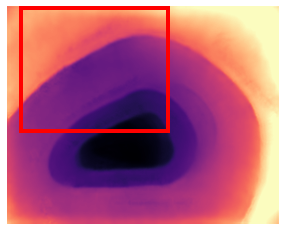} &
\includegraphics[height=\turnheightnew]{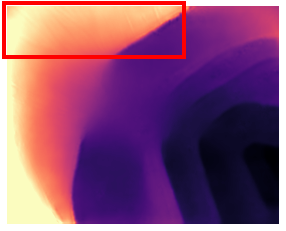} &
\includegraphics[height=\turnheightnew]{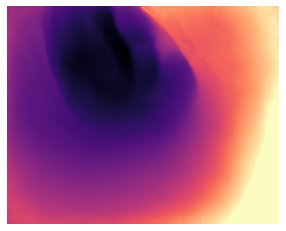}\\

 &
{\rotatebox{90}{\hspace{1mm}\scriptsize
{\bf Ours: ColDE}}} &
\includegraphics[height=\turnheightnew]{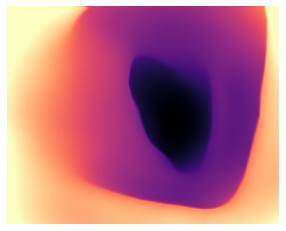} &
\includegraphics[height=\turnheightnew]{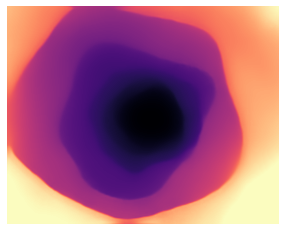} &
\includegraphics[height=\turnheightnew]{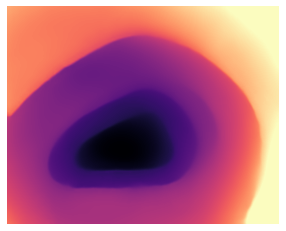} &
\includegraphics[height=\turnheightnew]{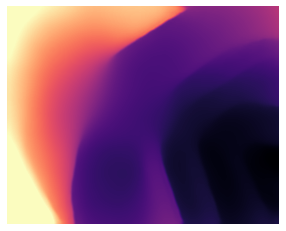} &
\includegraphics[height=\turnheightnew]{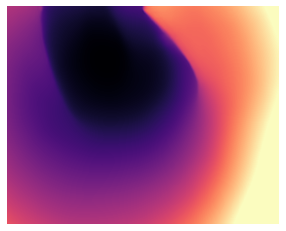}\\

\end{tabular}}
  \vspace{-3pt}
  \caption{Visual comparison of predicted depth maps on colon simulator data (more examples see the supplementary). The predictions from our model (last row) are smooth on the surface and sharp on the edge, resembling the ground-truth. Meanwhile, other methods are vulnerable to visual distractions, such as blood vessels and specular noise, predicting defective depth on these regions (circled in red).}
  \vspace{-3pt}
  \label{fig:sim_sota}
\end{figure*}

In Table~\ref{tab:sim_sota} we list the depth prediction results of our \textbf{ColDE} framework and other state-of-the-art monocular depth estimation methods.
For a fair comparison, only the self-supervised training objectives of these approaches were applied, and the trainings were conducted using the same training split of the simulator data described above.
When evaluated, the depth prediction of each frame was re-scaled to the median ground-truth, as in \cite{monodepth2}.

The results on the colonoscopy data show that our framework outperforms previous state-of-the-art methods that had been designed to give high accuracy for outdoor driving.
Due to our real-time requirement,
we also measured the execution speed of each approach.
When running on a single Nvidia Quadro RTX5000 GPU, our network generated depth maps at $36$ frames-per-second (fps), well faster than the colonoscopy video speed of about $20$ fps.

To visually compare the depth quality, in Fig.~\ref{fig:sim_sota} we show the depth map predictions from our method and from several previous works.
Overall, our approach produced smooth depths on the colon surface while maintaining sharp boundaries around haustral folds, resembling the colon shape in real-world and ground-truth depth maps.
Moreover, the other methods were vulnerable to visual distractions.
For example in the third case of Fig.~\ref{fig:sim_sota}, their depths were discontinuous around the specular regions on the image's top-left, near the haustral folds.
In the fourth case, the blood vessel texture on the top-left part of the image distracted the networks \cite{monodepth2,lyu2020hr}, resulting in unrealistic depth fluctuations.

\subsubsection{Ablation Study}
\begin{table*}[ht]
    \centering
    \begin{tabular}{cl|cccc|ccc|c}
        \multicolumn{2}{c}{\multirow{2}{*}{\bf Ablation}} & \multicolumn{4}{|c}{\bf Error metric $\downarrow$} & \multicolumn{3}{|c|}{\bf Accuracy metric $\uparrow$} & \\
        \cmidrule(lr){3-9}
         & & Abs Rel & Sq Rel & RMSE & RMSE log & $\delta < 1.25$ & $\delta < 1.25^2$ & $\delta < 1.25^3$ & \# \\
        \midrule
        \multirow{3}{*}{\rotatebox{90}{\bf Masks}} & Baseline ($M$) & 0.087 & 0.094 & 0.745 & 0.144 & 0.924 & 0.972 & 0.987 & 1 \\
        & $-M_{spec}$ & \underline{0.085} & \underline{0.091} & \underline{0.735} & \underline{0.142} & \underline{0.928} & \underline{0.973} & \underline{0.988} & 2 \\
        & $-M_{spec, valid}$ & 0.093 & 0.099 & 0.764 & 0.149 & 0.918 & 0.971 & \underline{0.988} & 3 \\
        \midrule
        \multirow{4}{*}{\rotatebox{90}{\bf Losses}} & Baseline ($L_{photo}$) & 0.087 & 0.094 & 0.745 & 0.144 & 0.924 & 0.972 & 0.987 & 4 \\
        & $+ L_{feat}$ & 0.085 & 0.091 & 0.737 & 0.143 & 0.925 & 0.971 & 0.987 & 5 \\
        & $+ L_{feat, depth}$ & 0.082 & 0.084 & 0.710 & 0.138 & 0.927 & 0.973 & 0.988 & 6 \\
        & $+ L_{feat, depth, norm}$ & \bf 0.077 & \bf 0.079 & \bf 0.701 & \bf 0.134 & \bf 0.935 & \bf 0.975 & \bf 0.989 & 7 \\
    \end{tabular}
    \caption{Ablation study. The effect of different masking schemes is shown in the first section. Although the specular mask improves depth quality visually, it has little impact on errors and accuracy; while the valid mask significantly improves performance. The impact of training objectives is shown in the second section. Our additional feature consistency loss and geometric losses can improve the depth quality.}
    \label{tab:sim_ablation}
\end{table*}

We study the effect of every mask and loss component in our framework.
Starting with the baseline model that is trained with only the appearance similarity objective $L_{photo}$ but with all masks $M$ applied, we gradually applied more objectives or eliminated mask components and studied the performance;
the results are shown in Table~\ref{tab:sim_ablation}.

\paragraph{Effect of masks}
As the specular mask is primarily to eliminate the effect of specular noise that is more common in optical colonoscopy~\cite{ma2021rnnslam},
we found that $M_{spec}$ did not show benefit on error or accuracy improvement on the simulator dataset (comparing line $2$ to $1$ in Table~\ref{tab:sim_ablation}).
Meanwhile, the valid mask is proven to be beneficial, as the errors significantly increased and accuracy decreased when we removed $M_{valid}$ from training between line $2$ and $3$.

\paragraph{Effect of objectives}
In the second part of Table~\ref{tab:sim_ablation}, we analyzed the effect of different loss functions.
After adding the feature consistency objective into the photometric losses, line $5$ shows improved depth quality.
The biggest improvement is provided by the geometric constraints.
In line $6$ and $7$ the errors drop significantly when $L_{depth}$ is added, and the accuracy has a substantial improvement after $L_{norm}$ is used.
Although the primary purpose of including geometric objectives is to improve the depth consistency between frames, they strengthen the network's ability to interpret shape information and thereby generate better depth maps for single images.

\subsection{Results on Optical Colonoscopy Videos}
To test our method on optical colonoscopy data, we selected videos from $85$ clinical procedures, where each video contains about $20$k frames.
From those videos we manually selected $79$ video clips in which the camera was in a moderate pulling back or pushing forward movement.
These testing sequences have $60$ to $300$ frames with an average of about $150$.
Additionally, we randomly sampled $185$k frames from the rest of the videos and used them and their adjacent frames to train the network; another $5$k-frame set was sampled for validation.

Testing predicted depth maps of optical colonoscopy data requires reconstructing 3D meshes.
We adopted the system in \cite{ma2021rnnslam}, where a SLAM component~\cite{engel2017direct} calculates corrected camera poses using the neural network's depth output and then a fusion component uses these poses to stitch the shapes of the frames together.
All the results reported below were generated with this system, but with the depth prediction network from the cited approaches.
The reconstruction system also provides a post-processing option named ``windowed depth averaging'' that averages the shapes of adjacent frames, eliminating outliers but sacrificing the real-time execution.
For comparison, we will show the reconstruction results with this post-processing step (with window size $7$ as in \cite{ma2021rnnslam}), along with the ones without.

\subsubsection{3D Reconstruction Visualization}
\begin{figure*}[!ht]
  \centering
  \resizebox{0.99\textwidth}{!}{
  \newcommand{\toprow}{0.195\columnwidth}
\newcommand{\secondrow}{0.25\columnwidth}
\newcommand{\avesecondrow}{0.27\columnwidth}

\centering

\begin{tabular}{@{\hskip 0mm}c@{\hskip 2mm}c@{\hskip 0mm}c@{\hskip 0mm}c@{\hskip 0mm}c@{}}

 &
RNNSLAM~\cite{ma2021rnnslam} &
Monodepth2~\cite{monodepth2} &
\textbf{ColDE} (-$L_{norm}$) &
\textbf{ColDE} (full)\\

\midrule

{\multirow{3}{*}[-2mm]{\rotatebox{90}{\large
{With averaging}}}} &
\includegraphics[height=\toprow]{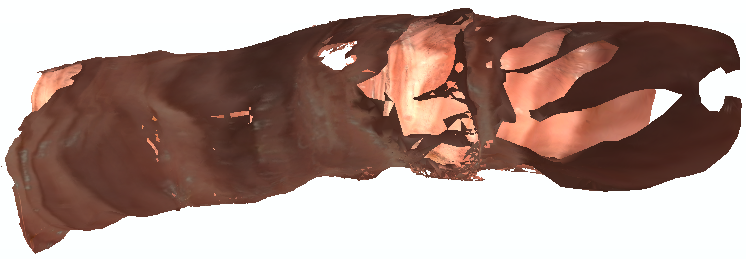} &
\includegraphics[height=\toprow]{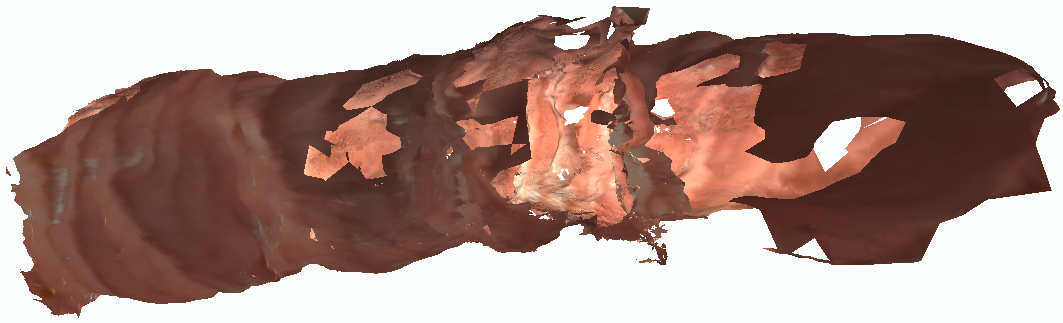} &
\includegraphics[height=\toprow]{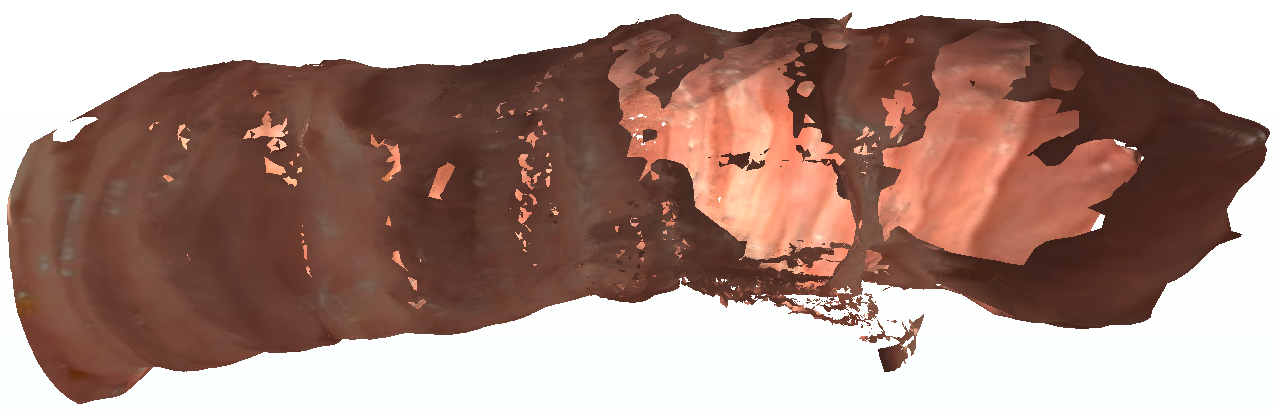} &
\includegraphics[height=\toprow]{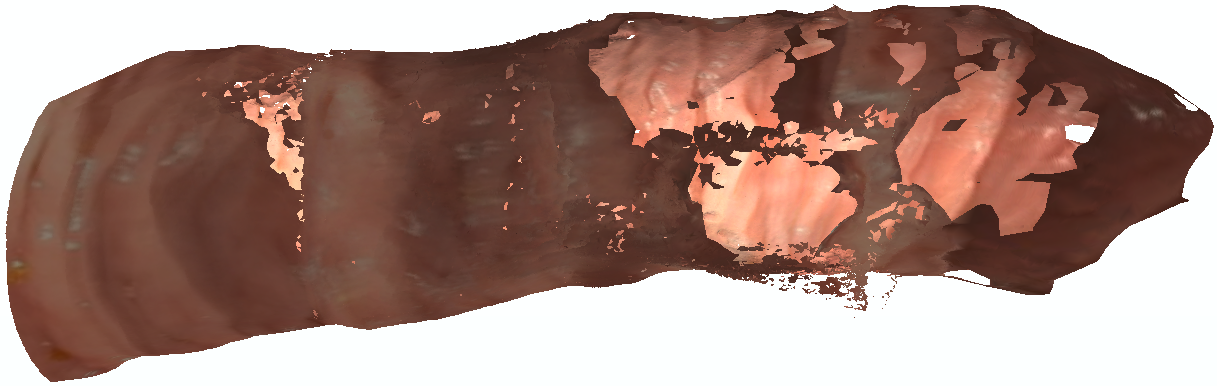} \\

&
\includegraphics[height=\avesecondrow]{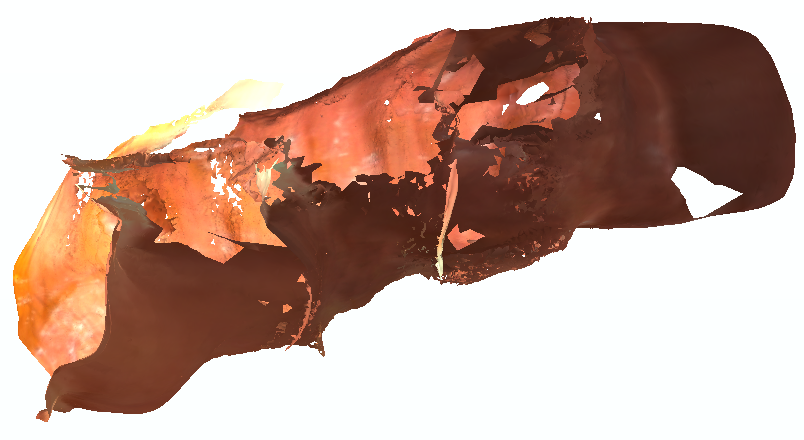} &
\includegraphics[height=\avesecondrow]{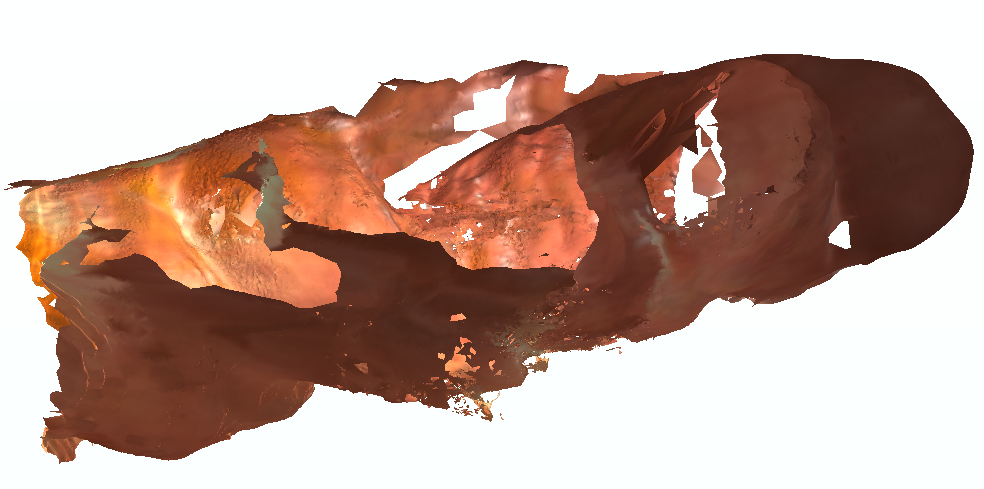} &
\includegraphics[height=\avesecondrow]{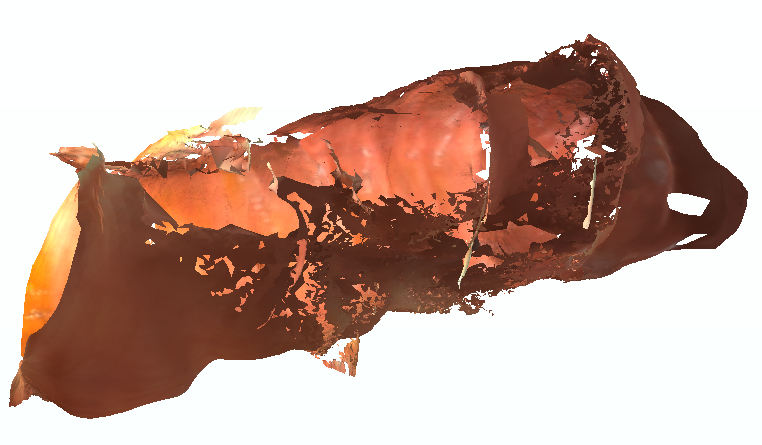} &
\includegraphics[height=\avesecondrow]{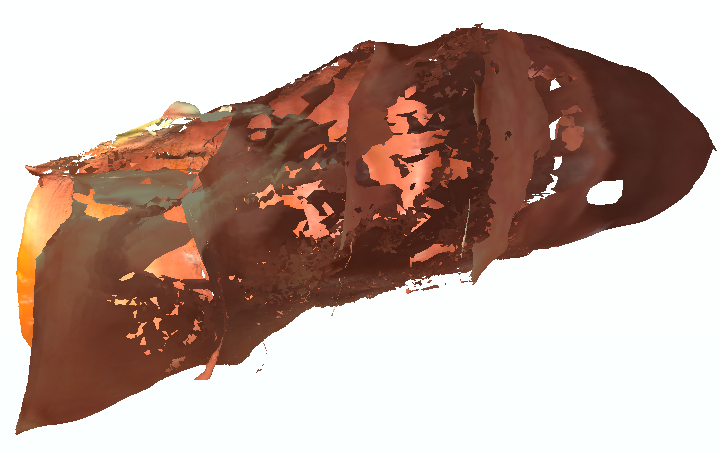} \\

&
\includegraphics[height=\secondrow]{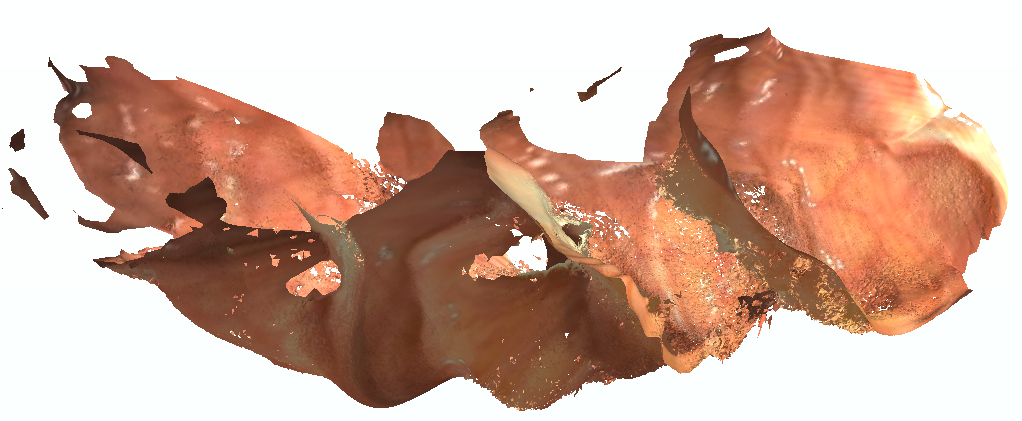} &
\includegraphics[height=\secondrow]{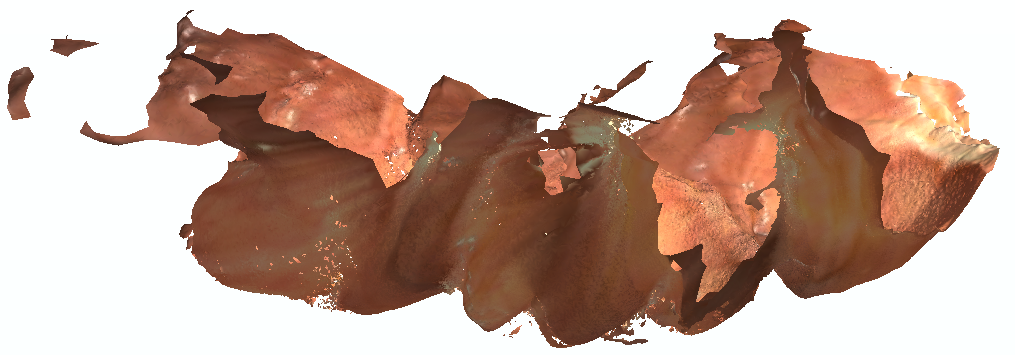} &
\includegraphics[height=\secondrow]{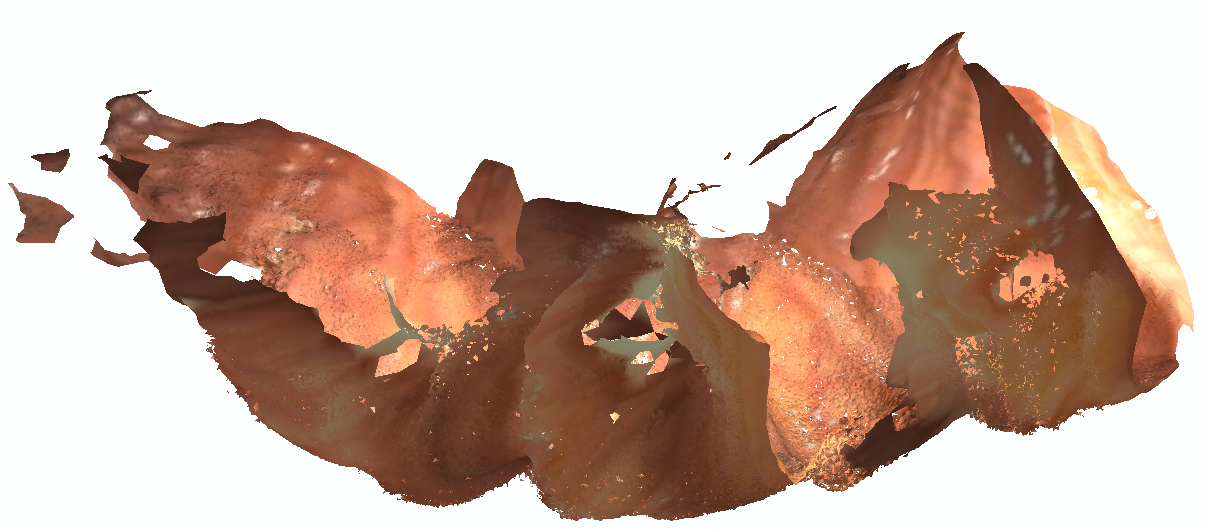} &
\includegraphics[height=\secondrow]{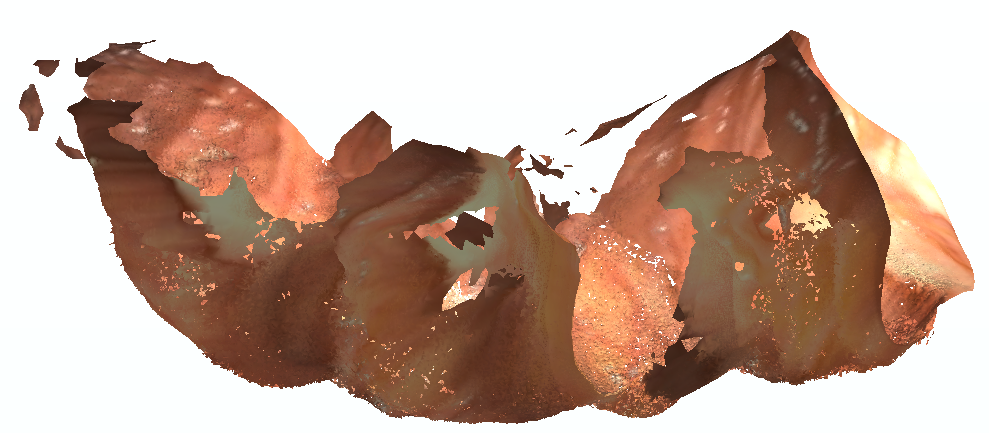} \\

\arrayrulecolor{black!30}\midrule

{\multirow{3}{*}{\rotatebox{90}{\large
{Without averaging}}}} &
\includegraphics[height=\toprow]{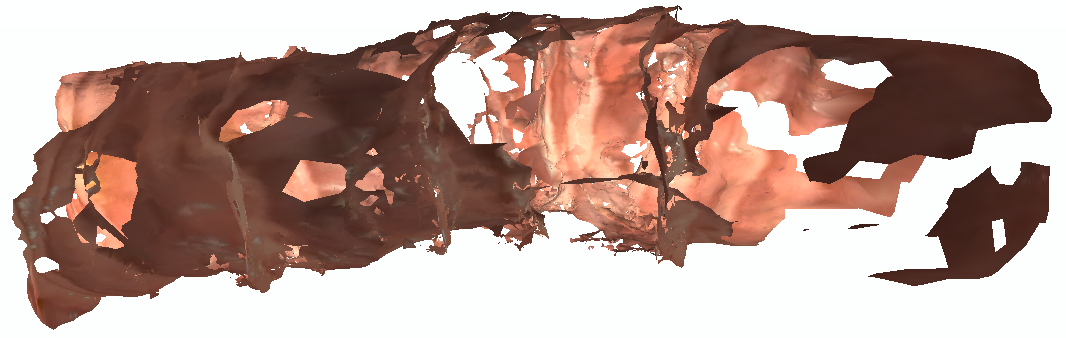} &
\includegraphics[height=\toprow]{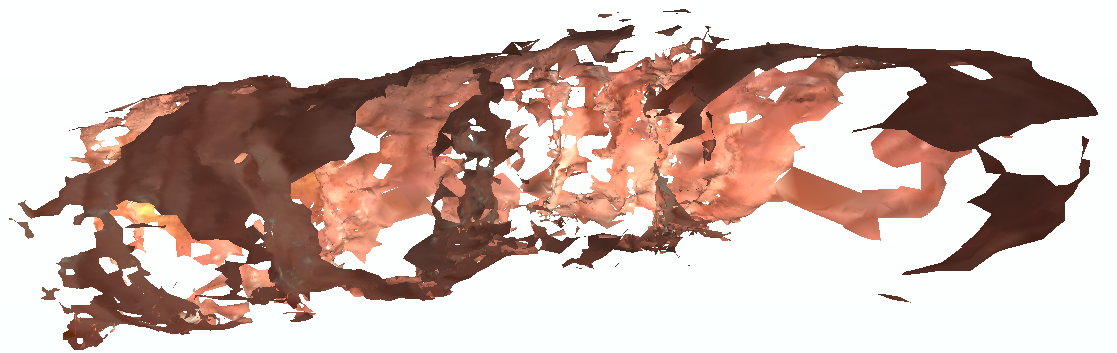} &
\includegraphics[height=\toprow]{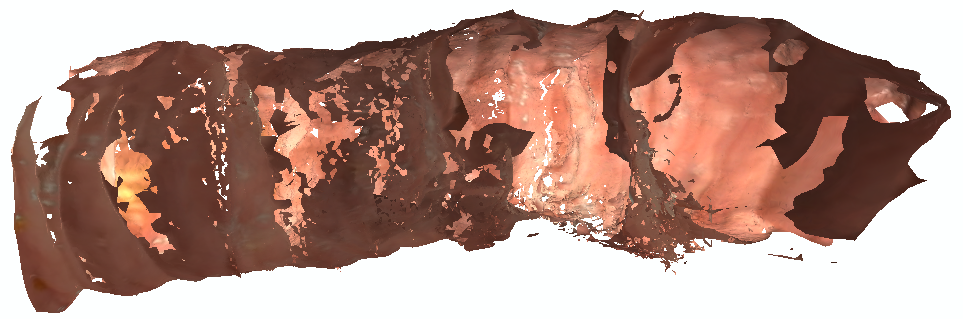} &
\includegraphics[height=\toprow]{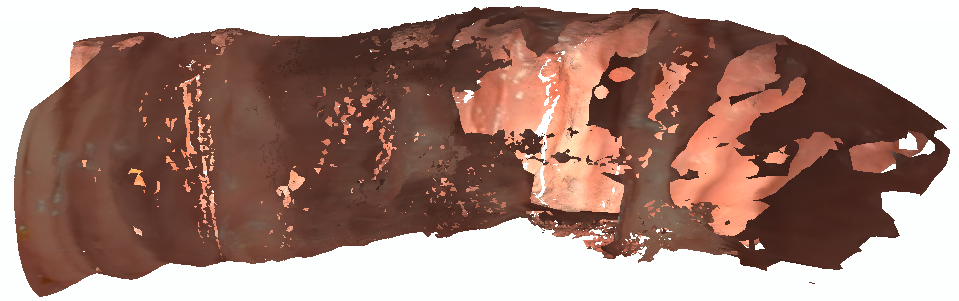} \\

&
\includegraphics[height=\secondrow]{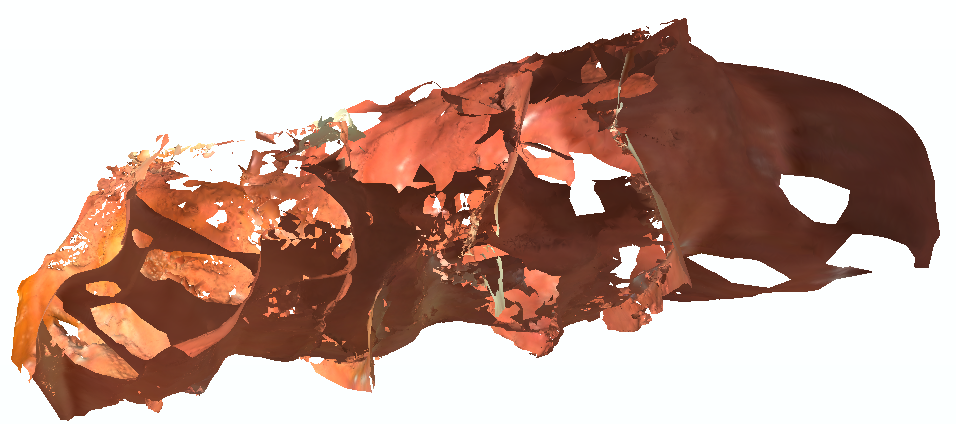} &
\includegraphics[height=\secondrow]{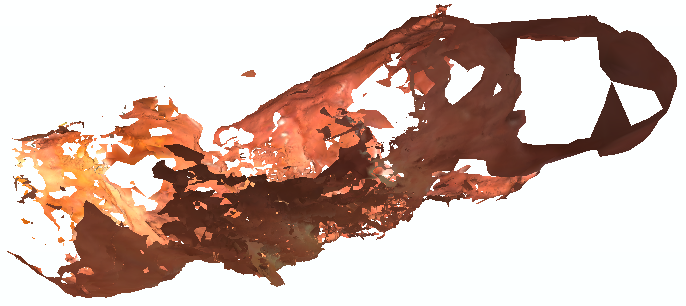} &
\includegraphics[height=\secondrow]{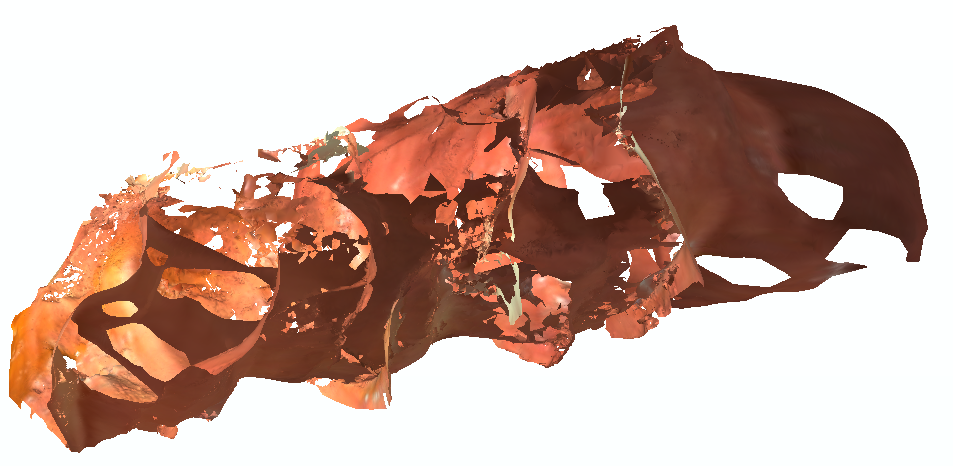} &
\includegraphics[height=\secondrow]{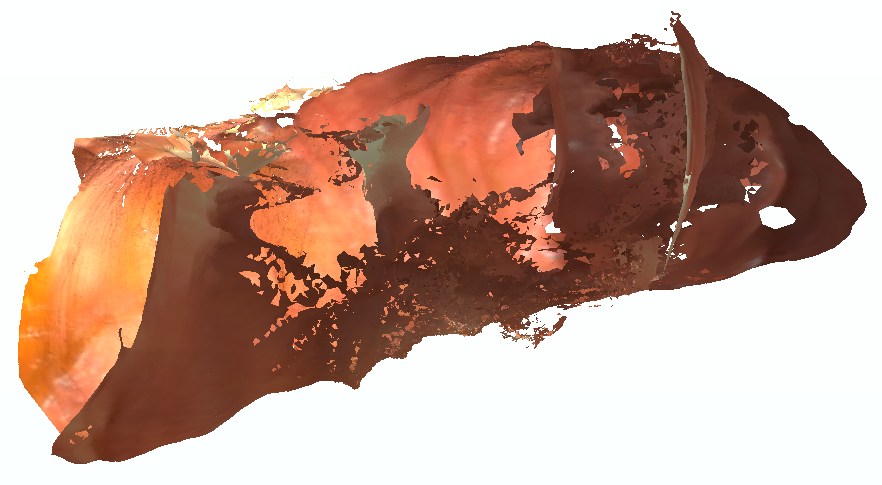} \\

&
\includegraphics[height=\secondrow]{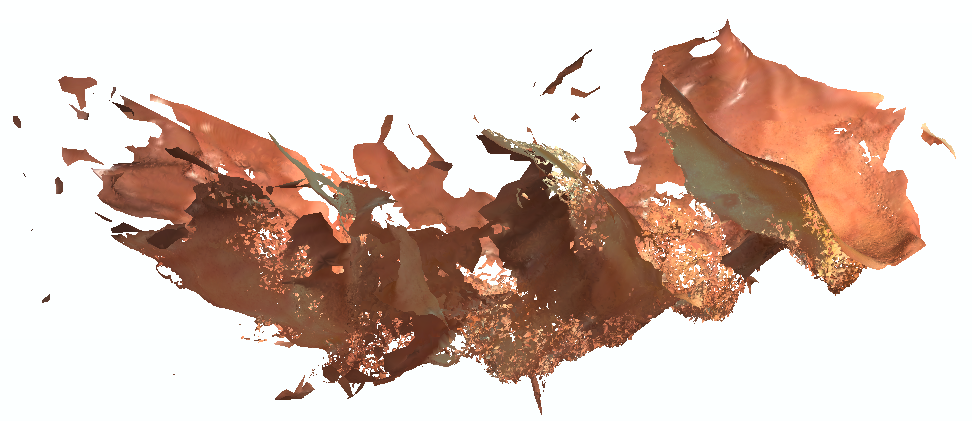} &
\includegraphics[height=\secondrow]{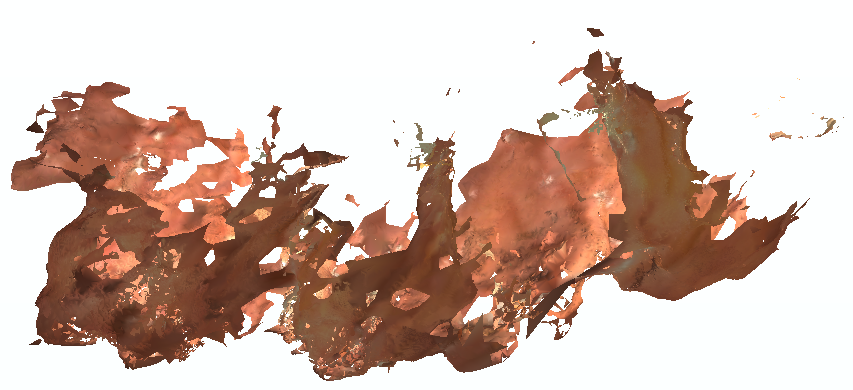} &
\includegraphics[height=\secondrow]{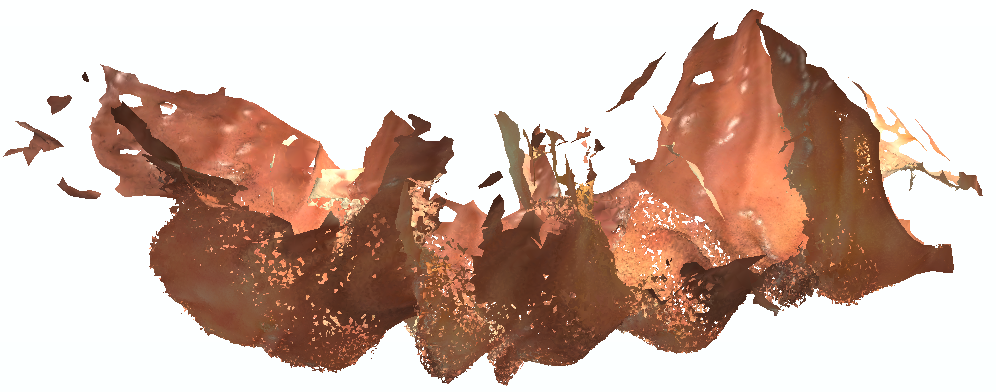} &
\includegraphics[height=\secondrow]{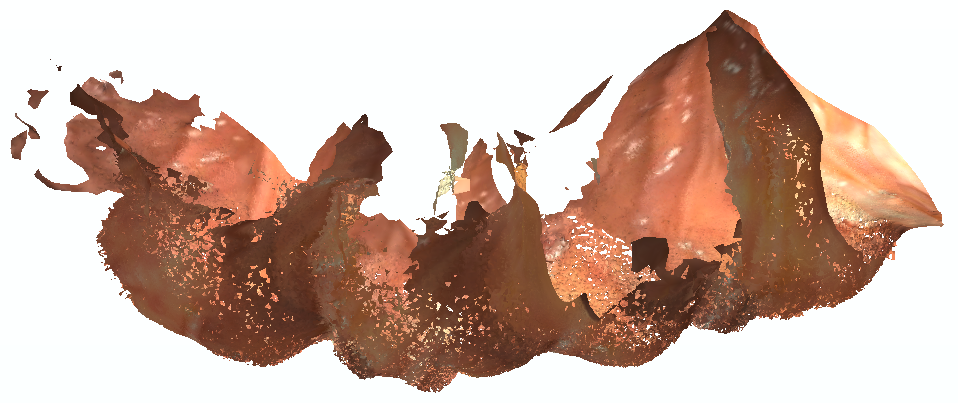} \\

\end{tabular}}
  \caption{3D reconstruction results on optical colonoscopy data. The meshes produced with ``windowed depth averaging''~\cite{ma2021rnnslam} are shown on the top, and the meshes produced without the averaging are on the bottom. Among the previous methods and the ablation, our full model is the only one that can produce good-quality meshes in real-time (without averaging).}
  \label{fig:real_sota}
  \vspace{-5pt}
\end{figure*}

Fig.~\ref{fig:real_sota} visualizes the reconstruction results of $3$ cases from our approach and previous works.
The results from \cite{ma2021rnnslam} were produced using their pre-trained network, which was trained with semi-supervision from real colonoscopy data, while the models of \cite{monodepth2} and \textbf{ColDE} were trained with our optical colonoscopy dataset.
Apart from the full model, we also show one ablation of our method that only included one geometric objective, corresponding to line $6$ in Table~\ref{tab:sim_ablation}.

Our model (the last column) is the only one to be able to reconstruct colon surfaces with good quality without using averaging post-processing; excluding the normal consistency loss deteriorates the result (penultimate column).
The common problem with other approaches is that their reconstructed surfaces are often sparse and broken due to shape misalignment among frames, thereby requiring averaging.
Moreover, when the shape information is not fully captured, artifacts with a ``skirt'' shape can occur around colon cylinders,
as seen in ``without averaging'' results in columns $1$ and $3$ of Fig.~\ref{fig:real_sota}.
However, these artifacts are less common in our full approach, as the network interprets shapes better when trained with geometric constraints.

\subsubsection{Expert Evaluation on Reconstruction Quality}
We conducted human evaluation of the reconstruction quality of $79$ testing sequences.
As instructed and verified by a colonoscopist, a member of our team judged the informativeness of the reconstructed meshes.
Reconstruction with the following characteristics is considered ideal:
1) the overall shape is similar to a generalized cylinder that resembles a human colon; 
2) the mesh contains few outliers and noise;
3) the surface has little unrealistic sparsity.
Based on the above standard, reconstruction quality was categorized into three categories, i.e., ``good'', ``moderate'' and ``poor''.

\begin{table}[ht]
    \centering
    \begin{tabular}{cl|ccc}
        \multicolumn{2}{c}{\multirow{2}{*}{\bf Methods}} & \multicolumn{3}{|c}{\bf Number of cases} \\
        \cmidrule(lr){3-5}
         & & Good & Moderate & Poor \\
        \midrule
        \midrule
        \multirow{3}{*}{\bf w/ ave} & RNNSLAM~\cite{ma2021rnnslam} & 44 & 27 & 8 \\
        & \textbf{ColDE} ($-L_{norm}$) & 57 & 16 & 6 \\
        & \textbf{ColDE} (full) & \bf 62 & 13 & \bf 4 \\
        \midrule
        \multirow{3}{*}{\bf w/o ave} & RNNSLAM~\cite{ma2021rnnslam} & 34 & 34 & 11 \\
        & \textbf{ColDE} ($-L_{norm}$) & 38 & 26 & 15 \\
        & \textbf{ColDE} (full) & \bf 62 & 13 & \bf 4 \\
    \end{tabular}
    \caption{Expert evaluation on the reconstruction quality. Our full model can produce more ``good'' meshes than the indicated alternatives.}
    \label{tab:real_sota}
    \vspace{-5pt}
\end{table}

As compared to the previous colonoscopy reconstruction approach~\cite{ma2021rnnslam} and our ablation,
the categorization results in Table~\ref{tab:real_sota} show that our framework produced notably more ``good'' reconstructions even without using windowed depth averaging.
Although the averaging post-processing can slightly change the sparsity of the mesh surface produced by our model as shown in Fig.~\ref{fig:real_sota}, it is seen to have little effect on the overall quality by category.
Importantly, it indicates that our method can capture the colon shape well enough to make the geometry post-processing unnecessary.

\subsection{Results on KITTI Dataset}
As detailed in the supplementary material, our method is competitive to the best of the published alternatives on the KITTI outdoor driving dataset~\cite{Geiger2012CVPR}.
\section{Conclusion}
Aiming to fundamentally improve the depth estimation quality of colonoscopy images, in this work we developed a set of training objectives to cope with their special challenges.
Geometric losses were designed to capture colon shape information and the photometric loss was extended to compensate noise.
Applied to colonoscopy video reconstruction, our network trained without any supervision is the first to be able to produce good-quality 3D meshes without post-processing, making the system clinically applicable.

\paragraph{Limitations and future work}
Although our approach brought significant improvement on colonoscopy depth estimation, there are still situations where the image illumination and texture are too complicated for our framework to handle.
For example, there is one category of our failures that we call the ``en face'' situation, where the camera directly faces the low-texture colon surface, confusing the current network.
To further improve the performance, optical flow may be useful to be included in the training, and the network architecture can be improved or re-designed to utilize temporal information.
We leave these further explorations to future work.

\paragraph{Acknowledgements}
We are grateful for the support from Shuai Zhang and his team for the use of their colonoscopy simulator.
We thank Olympus, Inc. for their funding and Zhen Li and his team at Olympus for their advice.

%%%%%%%%% REFERENCES
{\small
\bibliographystyle{ieee_fullname}
\bibliography{egbib}
}

\newpage
\onecolumn

\begin{center}
\textbf{\Large ColDE: A Depth Estimation Framework for Colonoscopy Reconstruction \vspace{3pt} \\
Supplementary Material}
\end{center}
\vspace{10pt}

\appendix

\section{Results on KITTI Dataset}
\begin{table*}[h!]
    \centering
    \begin{tabular}{lc|cccc|ccc}
        \multirow{2}{*}{\bf Methods} & \multirow{2}{*}{\bf Resolution} & \multicolumn{4}{c|}{\bf Error metric $\downarrow$} & \multicolumn{3}{c}{\bf Accuracy metric $\uparrow$} \\
        \cmidrule(lr){3-9}
         & & Abs Rel & Sq Rel & RMSE & RMSE log & $\delta < 1.25$ & $\delta < 1.25^2$ & $\delta < 1.25^3$\\
        \midrule
        SfMLearner~\cite{zhou2017unsupervised} & $416 \times 128$ & 0.208 & 1.768 & 6.856 & 0.283 & 0.678 & 0.885 & 0.957 \\
        Vid2Depth~\cite{mahjourian2018unsupervised} & $416 \times 128$ & 0.163 & 1.240 & 6.220 & 0.250 & 0.762 & 0.916 & 0.968 \\
        Struct2Depth~\cite{casser2019depth} & $416 \times 128$ & 0.141 & 1.026 & 5.291 & 0.215 & 0.816 & 0.945 & 0.979 \\
        SC-SfMLearner~\cite{bian2019unsupervised} & $832 \times 256$ & 0.137 & 1.089 & 5.439 & 0.217 & 0.830 & 0.942 & 0.975 \\
        Monodepth2~\cite{monodepth2} & $640 \times 192$ & 0.115 & 0.903 & 4.863 & 0.193 & 0.877 & 0.959 & 0.981 \\
        PackNet-SfM~\cite{packnet} & $640 \times 192$ & 0.111 & \bf 0.785 & \bf 4.601 & 0.189 & 0.878 & 0.960 & 0.982 \\
        HR-Depth~\cite{lyu2020hr} & $640 \times 192$ & \bf 0.109 & 0.792 & 4.632 & \bf 0.185 & \bf 0.884 & \bf 0.962 & \bf 0.983 \\
        \midrule
        \textbf{ColDE} (baseline) & $640 \times 192$ & 0.120 & 0.946 & 4.937 & 0.197 & 0.869 & 0.958 & 0.980 \\
        \textbf{ColDE} (full) & $640 \times 192$ & 0.119 & 0.910 & 4.820 & 0.195 & 0.871 & 0.958 & 0.981
    \end{tabular}
    \caption{Results on KITTI. The listed results of the previous work are from their cited literature, all produced with the same settings: monocular inputs, self-supervision only, and KITTI training only. When testing our framework (minus the specular mask) on the outdoor driving data, its performance is comparable to Monodepth2.}
    \label{tab:kitti_sota}
\end{table*}

Although our framework is designed to handle the specific challenges of colonoscopy data, to further prove our method's validity, we also tested it on the widely used outdoor driving dataset of depth evaluation: KITTI~\cite{Geiger2012CVPR}.
Following the common practice we used the data split in \cite{eigen2015predicting} for training and testing, and excluded static frames~\cite{zhou2017unsupervised}.
The input images of our network are resized to $640 \times 192$.
The \textit{\textbf{ColDE} (full)} result listed in Table~\ref{tab:kitti_sota} was generated from our full framework but without implementing the specular mask $M_{spec}$;
our baseline model further excluded extra losses with only $L_{photo}$ applied.

Tested on the outdoor driving dataset, our method's performance is comparable to Monodepth2~\cite{monodepth2}, only below PackNet-SfM~\cite{packnet} and HR-Depth~\cite{lyu2020hr} who implemented more sophisticated network architectures that potentially slow down the execution (as shown in the experiment section of the main paper).
As the human-made world in KITTI contains mostly flat surfaces with simple geometry as well as the photometric noise appears infrequently when lit by sunlight, our geometric constraints and image feature loss designed for colonoscopy data brought moderate performance improvement compared to our baseline.

\newpage

\section{Additional Results: Simulator Depth Maps Comparison}
\begin{figure*}[!ht]
  \centering
  \resizebox{0.99\textwidth}{!}{
  \newcommand{\turnheightnew}{0.097\columnwidth}

\centering

\begin{tabular}{@{\hskip 0mm}c@{\hskip 1mm}c@{\hskip 0mm}c@{\hskip 0mm}c@{\hskip 0mm}c@{\hskip 0mm}c@{\hskip 0mm}c@{\hskip 0mm}c@{}}

{\rotatebox{90}{\hspace{6mm}\scriptsize{Input}}} &
\includegraphics[height=\turnheightnew]{figs/sim_sota/NLY1490_img.png} &
\includegraphics[height=\turnheightnew]{figs/sim_sota/NLY812_img.png} &
\includegraphics[height=\turnheightnew]{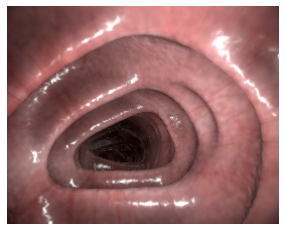} &
\includegraphics[height=\turnheightnew]{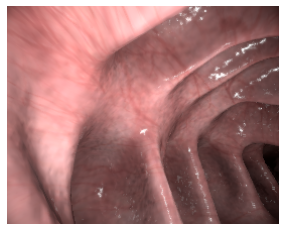} &
\includegraphics[height=\turnheightnew]{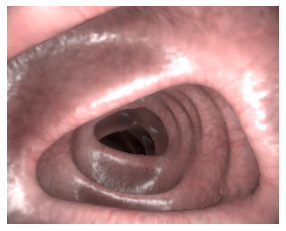} &
\includegraphics[height=\turnheightnew]{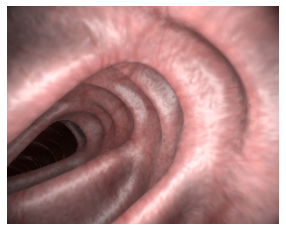} &
\includegraphics[height=\turnheightnew]{figs/sim_sota/CJM52_img.png}\\

{\rotatebox{90}{\hspace{1.5mm}\scriptsize
{Ground-truth}}} &
\includegraphics[height=\turnheightnew]{figs/sim_sota/NLY1490_gt.png} &
\includegraphics[height=\turnheightnew]{figs/sim_sota/NLY812_gt.png} &
\includegraphics[height=\turnheightnew]{figs/sim_sota/QGQ432_gt.png} &
\includegraphics[height=\turnheightnew]{figs/sim_sota/NLY912_gt.png} &
\includegraphics[height=\turnheightnew]{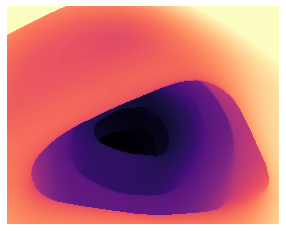} &
\includegraphics[height=\turnheightnew]{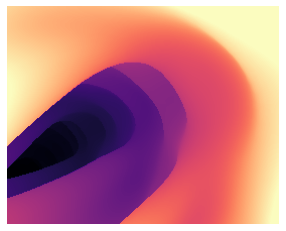} &
\includegraphics[height=\turnheightnew]{figs/sim_sota/CJM52_gt.png}\\

{\rotatebox{90}{\hspace{2.6mm}\tiny
{SfMLearner~\cite{zhou2017unsupervised}}}} &
\includegraphics[height=\turnheightnew]{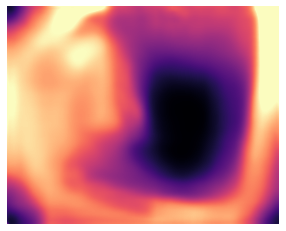} &
\includegraphics[height=\turnheightnew]{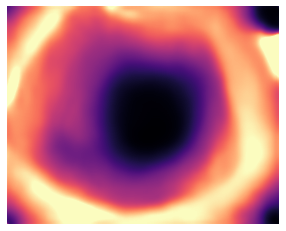} &
\includegraphics[height=\turnheightnew]{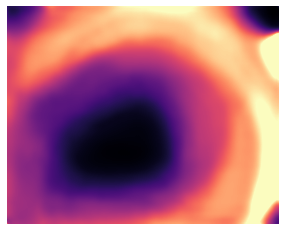} &
\includegraphics[height=\turnheightnew]{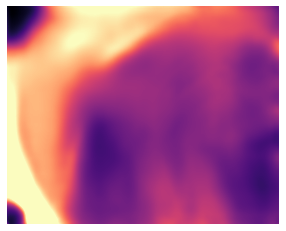} &
\includegraphics[height=\turnheightnew]{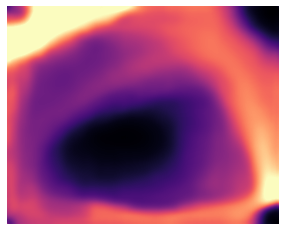} &
\includegraphics[height=\turnheightnew]{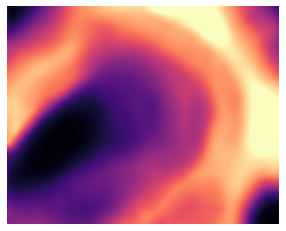} &
\includegraphics[height=\turnheightnew]{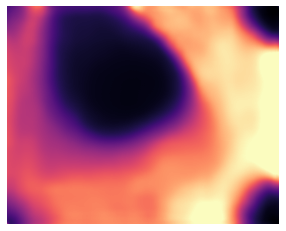}\\

{\rotatebox{90}{\hspace{1.5mm}\tiny
{SC-SfMLearner~\cite{bian2019unsupervised}}}} &
\includegraphics[height=\turnheightnew]{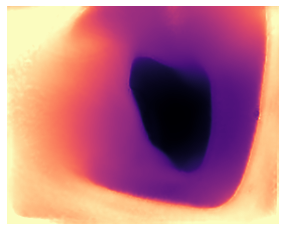} &
\includegraphics[height=\turnheightnew]{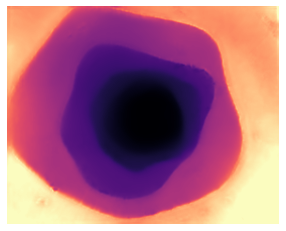} &
\includegraphics[height=\turnheightnew]{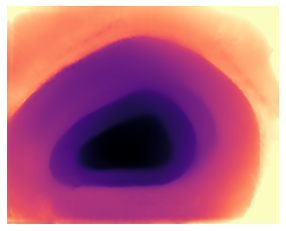} &
\includegraphics[height=\turnheightnew]{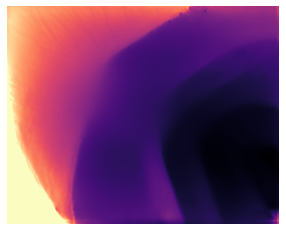} &
\includegraphics[height=\turnheightnew]{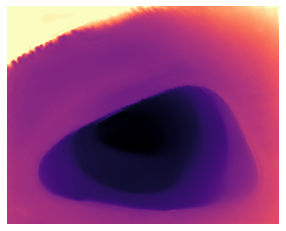} &
\includegraphics[height=\turnheightnew]{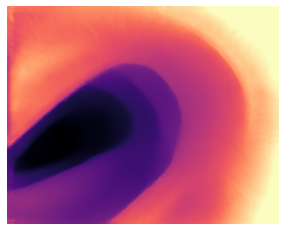} &
\includegraphics[height=\turnheightnew]{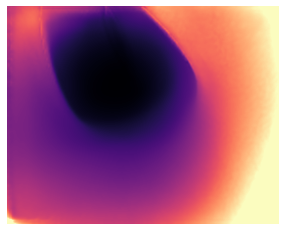}\\

{\rotatebox{90}{\hspace{2.5mm}\tiny
{Monodepth2~\cite{monodepth2}}}} &
\includegraphics[height=\turnheightnew]{figs/sim_sota/NLY1490_mono.png} &
\includegraphics[height=\turnheightnew]{figs/sim_sota/NLY812_mono.png} &
\includegraphics[height=\turnheightnew]{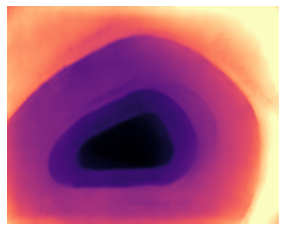} &
\includegraphics[height=\turnheightnew]{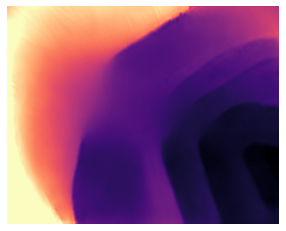} &
\includegraphics[height=\turnheightnew]{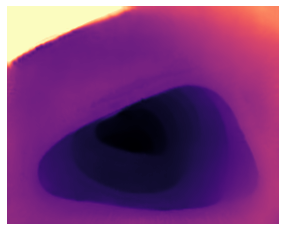} &
\includegraphics[height=\turnheightnew]{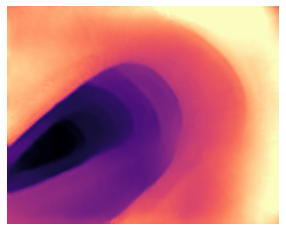} &
\includegraphics[height=\turnheightnew]{figs/sim_sota/CJM52_mono.png}\\

{\rotatebox{90}{\hspace{2mm}\tiny
{PackNet-SfM~\cite{packnet}}}} &
\includegraphics[height=\turnheightnew]{figs/sim_sota/NLY1490_packnet.png} &
\includegraphics[height=\turnheightnew]{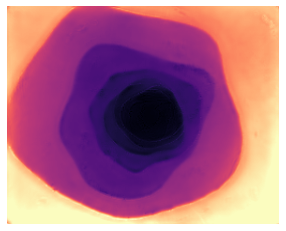} &
\includegraphics[height=\turnheightnew]{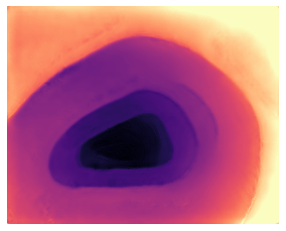} &
\includegraphics[height=\turnheightnew]{figs/sim_sota/NLY912_packnet.png} &
\includegraphics[height=\turnheightnew]{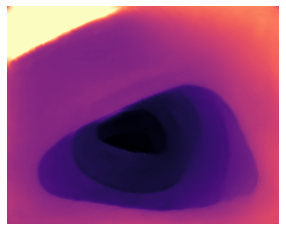} &
\includegraphics[height=\turnheightnew]{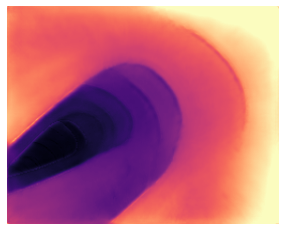} &
\includegraphics[height=\turnheightnew]{figs/sim_sota/CJM52_packnet.png}\\

{\rotatebox{90}{\hspace{3mm}\tiny
{HR-Depth~\cite{lyu2020hr}}}} &
\includegraphics[height=\turnheightnew]{figs/sim_sota/NLY1490_hrdepth.png} &
\includegraphics[height=\turnheightnew]{figs/sim_sota/NLY812_hrdepth.png} &
\includegraphics[height=\turnheightnew]{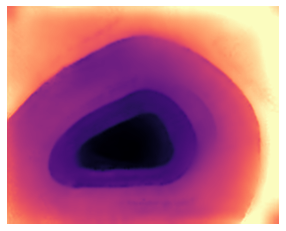} &
\includegraphics[height=\turnheightnew]{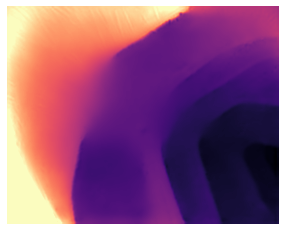} &
\includegraphics[height=\turnheightnew]{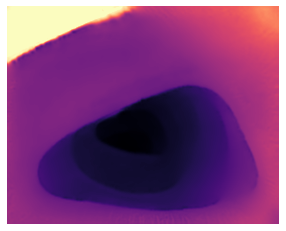} &
\includegraphics[height=\turnheightnew]{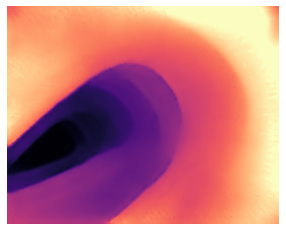} &
\includegraphics[height=\turnheightnew]{figs/sim_sota/CJM52_hrdepth.png}\\

{\rotatebox{90}{\hspace{1mm}\scriptsize
{\bf Ours: ColDE}}} &
\includegraphics[height=\turnheightnew]{figs/sim_sota/NLY1490_ours.png} &
\includegraphics[height=\turnheightnew]{figs/sim_sota/NLY812_ours.png} &
\includegraphics[height=\turnheightnew]{figs/sim_sota/QGQ432_ours.png} &
\includegraphics[height=\turnheightnew]{figs/sim_sota/NLY912_ours.png} &
\includegraphics[height=\turnheightnew]{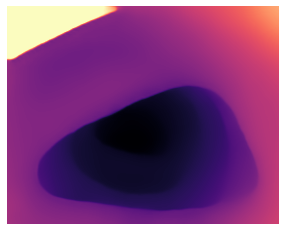} &
\includegraphics[height=\turnheightnew]{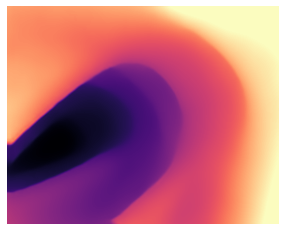} &
\includegraphics[height=\turnheightnew]{figs/sim_sota/CJM52_ours.png}\\

\end{tabular}}
  \vspace{-3pt}
  \caption{Additional visual comparison of predicted depth maps on colon simulator data. Compared to all the previous work, our model (last row) produced the best-quality depth maps that resemble the geometry of colon surfaces.}
  \label{fig:sim_sota_supp}
\end{figure*}

\end{document}